\documentclass[a4paper,
preprint,
aps,pra,superscriptaddress,floatfix,showpacs,showkeys]{revtex4-1}

\usepackage[english]{babel}
\usepackage{amsmath}
\usepackage{amssymb,graphicx}
\usepackage{braket}
\usepackage{ifpdf,color,tabularx,multirow}

\usepackage{color}
\definecolor{darkblue}{rgb}{0,0,.6}
\definecolor{darkgreen}{rgb}{0,0.5,0}

\usepackage{ifthen}

\definecolor{darkblue}{rgb}{0,0,.6}

\ifpdf
  \usepackage{epstopdf}
  \usepackage[pdftex,unicode,
  pdfstartview={FitH},pdfborder={0 0 0}]{hyperref}
  \usepackage{hypcap}
\else
  \usepackage[hypertex]{hyperref}
\fi
\hypersetup{
  bookmarksnumbered = true,
  colorlinks = true, linkcolor = darkblue,
  citecolor = darkblue, filecolor = darkblue,
  menucolor = darkblue, urlcolor = darkblue
}

\usepackage{bm}

\newcolumntype{C}[1]{>{\centering\arraybackslash}m{#1}}

\DeclareMathAlphabet \mathbfcal{OMS}{cmsy}{b}{n}
\usepackage{physics}

\graphicspath{ {Figures/} }

\begin{document}
\title{Ultrafast and Strong-Field Physics in Graphene-Like Crystals: Bloch Band Topology and High-Harmonic Generation}

\author{Hamed Koochaki Kelardeh}
\email[]{hkelardeh@pks.mpg.de}
\affiliation{Max Planck Institute f\"ur Physik komplexer Systems, N\"othnitzer Str. 38, 01187 Dresden, Germany}
\begin{abstract}

The emerging possibilities to steer and control electronic motion on subcycle time scales with strong electric fields enable studying the nonperturbative optical response and Bloch bands' topological properties, originated from Berry's trilogy: connection, curvature, and phase.
This letter introduces a theoretical framework for the nonperturbative electron dynamics in two-dimensional (2D) crystalline solids induced by the few-cycle and strong-field optical lasers. 
In the presented model, the expression associated with the Bloch band topology and broken crystal symmetry merges self-consistently in the system observables such as High Harmonic Generation (HHG). 
 This singles out our work from recent HHG calculations from the strongly-driven systems. 

Concisely, in our theoretical experiment on 2D materials in the strong-field optical regime, we show that Bloch band topology and broken symmetry manifest themselves in several ways: 
the momentum-resolved attosecond interferometry of electron wave packets, anomalous and chiral velocity in both intraband and interband dynamics, anomalous Hall current and respective HHG highly sensitive to the laser waveform, multiple plateau-cutoff structures in both longitudinal and transverse HHG, the formation of even harmonics in the perpendicular polarization with respect to the driving laser, singular jumps across the phase diagram of the HHG, attosecond chirp, and ultrafast valley polarization induced by the chiral gauge field that is robust to lattice imperfections and scattering. 
The link between HHG and solid-state band geometry offers an all-optical reconstruction of electron band structure by optical means, and accelerates studies on the non-equilibrium Floquet engineering, topologically-protected nonlinear spin and edge currents, valleytronics, quantum computing and high-temperature superconductivity on sub-femtosecond time scales.

\end{abstract}


\maketitle

\section{Introduction}

Today, geometric effects in controlling various phases of matter – from superconducting to semimetallic to topological insulating with conducting edge or surface states - is a central topic in condensed matter physics \cite{Xiao_Berry_2010}. 
In adiabatic processes, the geometric phase induces such effects as the quantum Hall \cite{Zhang_Experimental_2005}. However, the importance of geometric effects in nonadiabatic processes has been largely overlooked, up until very recently \cite{Silva2019,Chacon2020,Gaarde_PRL2020,Bauer2020,Avetissian_HHG_2020}. Such nonadiabatic effects become particularly important for nonperturbative nonlinear optical phenomena. 

Recent advancement in strong-field and ultrafast laser technology has brought  nonlinear optics and condensed matter physics into a new perspective.  Processes like  dynamical Franz-Keldysh effect \cite{Lucchini_Franz_Keldysh_2016,Yabana_Franz_Keldysh_2016}, optical Faraday rotation \cite{Wismer_Faraday_2017}, Landau–Zener tunneling \cite{Kitamura_Landau_Zener_2020} have proposed and experimentally probed in solids by means of attosecond science.
Among various nonperturbative response of matter in the strong-field regime, high-harmonic generation (HHG) has drawn significant attention in the solid-state community [started with \cite{Ghimire_2011}].
The advantage of HHG over conventional spectroscopic methods such as angle-resolved photoemission spectroscopy, photogalvanic effect, and Kerr rotation is the possibility to achieve sub-cycle temporal resolution.
High-order harmonic spectra, undeniably, contains rich information about the band structure \cite{Lanin2017}, Berry curvature \cite{Luu_Worner_2018}, topology, and phase transition of solid materials \cite{Silva2019,Chacon2020}.
Representative applications of solid-state HHG include generating coherent table-top extreme ultraviolet (XUV) \cite{Goulielmakis_2015} and  attosecond pulses \cite{Garg_Goulielmakis_2018}, sub-cycle recollision dynamics of electron-hole pairs \cite{SchubertO_2014,Hohenleutner_2015,McDonald_PRA_2015}, and all-optical retrieval of electronic band structures \cite{Vampa_PRL_2015,Zhao_All_optical_2019}.

Amongst various class of solid-state systems, 2D materials establish a remarkable platform to investigate the ultrafast and strong optical phenomena due to their tunability, integrability, flexibility, and high quantum efficency governed by their direct energy gap \cite{Xu_two_dimensional_2013}.  
There is a broad range of novel 2D materials and van der Waals heterostructures \cite{Novoselov_2D_material_2016} where the time-reversal is preserved. Still, the spatial symmetry is intrinsically broken, and non-zero Berry curvature at ${\bf{K}}$ and ${\bf{K'}}$ valleys is generated . Subsequently, the nontrivial Berry curvature leads to a quantum anomalous Hall effect \cite{Weng_Quantum_anomalous_2015,Rubio2019,Cavalleri2020,Stockman_Anomalous_2020}.

To date, interpreting the attosecond response of crystal driven by an intense light field has mostly focused on the role of the band structure. The role of geometrical properties in this context has been explored to a very limited extent. Indeed, new HHG selection rules in the strong-field regime are mandated to uncover nonadiabatic geometric effects, ultrafast chirality and dynamical symmetry breaking spectroscopy.
In this letter, we present a gauge-invariant theory of strong-field dynamics that enables the characterization of crystal symmetry in electronic properties and the Bloch band topology of honeycomb 2D materials. Our model goes beyond the semiclassical Boltzmann theory and expands the Liouville von Neumann equation and solves the "generalized" reduced density operator for the dissipative system interacting with the nonperturbative gauge field. Hence, it takes the impact of dephasing and decoherence on the system observables into account. 

We extend the time-dependent density matrix formalism and highlight the proportion of Berry's trilogy $-$ connection, curvature and phase $-$ on carrier trajectory and momentum-resolved population distribution, chirality and pseudospin textures of massless $|$ massive Dirac fermions, intraband and interband anomalous velocity, and more notably, on high harmonic emission from longitudinal (drift) and transverse (Hall) currents. 
The Berry-assisted terms in the photoemission spectra predominantly participate in the above-bandgap harmonic where the intraband excitations are seemingly dominating the HHG spectra.

The electron population distribution in the reciprocal space shows an interference pattern due to the rapid phase modulation of the electron wave packet near the Dirac cones. 
The relative strength of interband and intraband anomalous velocity indicates the crossover between the multiphoton excitation and tunneling regime. 
Specifically, for our hexagonal boron nitride (hBN) system under scrutiny, the combination of laser parameters versus material properties governs a carrier-wave Rabi flopping (CWRF) process \cite{Yakovlev_Colloquium_2018}. The observation of CWRF indicates a strong correlation between interband transition (Rabi oscillation) and intraband motion (Bloch oscillations) anticipated in the intermediate Keldysh regime.

The HHG time profile exhibits bursts of emission where their attosecond chirp differs from the gaseous HHG \cite{Krausz_Ivanov_2009}.
Additionally, the phase of the high-harmonic spectra delivers rich information about the multiple plateau behavior of sold HHG. The sharp jumps in the spectral phase may also use as highly sensitive machinery to characterize topological phase transition in the nontrivial class of materials such as Chern, $\mathbb{Z}_2$, and crystalline topological insulators, as well as Weyl semimetal.

Furthermore, we postulate that topologically-protected valley polarization universally takes place under chiral gauge-filed in graphene-like nanocrystals regardless of their fine lattice structures and chemical composition, spin-orbit coupling strength.
Indeed, the emergence of ultrafast and topological photonics with topological material will continue to fascinate condensed matter physics, nonlinear optics and attosecond communities and pave the route to quantum computing and high-temperature topological superconductor.

\section{Methods}
\label{sec:method}

\begin{figure}
	\begin{center}
		\includegraphics[width=\linewidth]{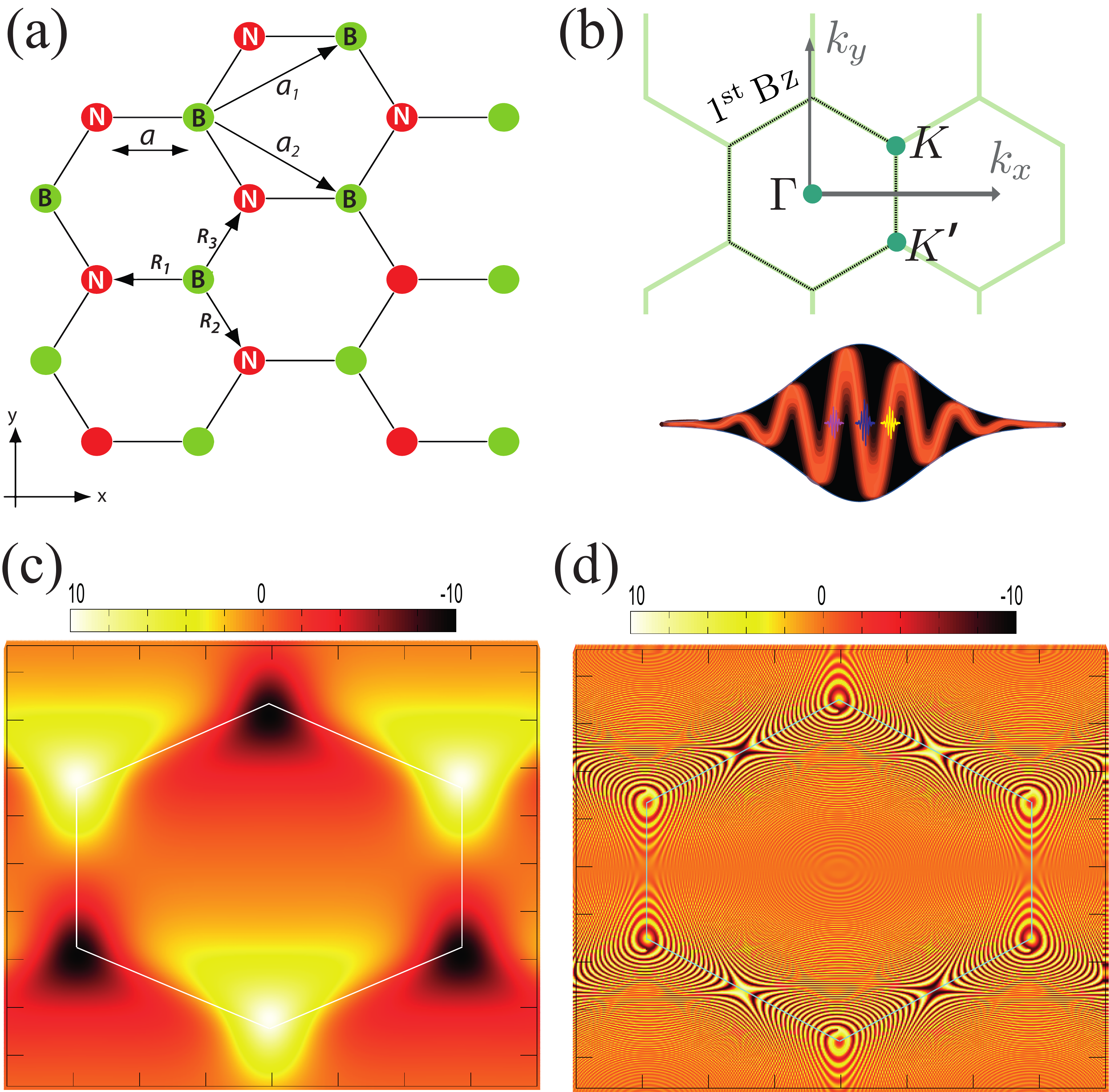}
		\caption{\label{Fig_schematic} (a) configuration space and (b) reciprocal space band structure of hexagonal Boron Nitride. The lattice constant, unit vectors, and three nearest-neighbor vectors are labeled. The incident laser waveform and the created  higher frequency signals are schematically depicted. (c) Intraband and (d) interband velocities illustrated in the reciprocal space using Eq. \ref{Eq:velocity_inter_intra}. The trigonal wrapping, chirality and lightwave-induced anomalous Hall effect originating from the band topology is evidently observed. }
	\end{center}
\end{figure}

We form a computational method to describe nonlinear electron dynamics in two dimensional (2D) crystalline solids induced by the few-cycle strong-field optical lasers. 
We have applied our model to monolayer hexagonal boron nitride (hBN) as a prototypical system; however, it can be equivalently applied to other 2D gapped systems such as topological crystalline insulators, Transition metal dichalcogenides (TMDs), and gapped graphene. Such crystals allow the simultaneous breakage of time-reversal symmetry and inversion symmetry. The mathematical justification of such a universality argument is given in Supplementary Information [\hyperlink{SMSpinor}{S1}].
The field-free Hamiltonian for the Bloch electrons in the $\pi $-bands of hBN reads
\begin{equation}
	\label{Eq:Hamiltonian_hBN}
	{{\cal H}_0} = \left( {\begin{array}{*{20}{c}}
			{{E_{\rm{B}}}}&{g({\bf{q}})}\\
			{{g^*}({\bf{q}})}&{{E_{\rm{N}}}}
	\end{array}} \right)
\end{equation}
$E_{\rm{B}}$ and $E_{\rm{N}}$ are the energy at the boron and nitrogen site,  respectively. 
$g({\bf{q}})$ is a complex function of the sum of the phase factors ${e^{i{\bf{k}} \cdot {{\bf{R}}_i}}}$ with nearest-neighboring vectors ${{\bf{R}}_i}$ $(i = 1,...,3)$ [Fig. \ref{Fig_schematic} (a) and (b)]. 
The wave vector in the Brillouin zone is written as $\mathbf{q} = ({q_x},{q_y})$ and the eigenvalues of Hamiltonian (\ref{Eq:Hamiltonian_hBN}) are given
by ${E_{c,v}}({\bf{q}}) = {E_0} \pm \sqrt {{m^2} + {{\left| {g({\bf{q}})} \right|}^2}} $
where ${E_0} = ({E_B} + {E_N})/2$ and $m = ({E_B} - {E_N})/2$ is the energy gap. The $ + $ and $-$ signs correspond to the conduction (c) and valence (v) band, respectively. The tight-binding hopping parameters and on-site energies are obtained from the ab-initio calculation\cite{Ribeiro_2011}.

An applied electric field generates both the intraband (adiabatic) and interband (nonadiabatic) electron dynamics. The intraband dynamics is determined by the Bloch acceleration theorem in the reciprocal space.  
For an electron with initial momentum $\bf{q}$ the electron dynamics is described by the time-dependent wave vector given by ${\bf{k}}(t) = {\bf{q}} - e{\hbar ^{ - 1}}{\bf{A}}(t)$ with  ${\bf{A}}(t) =  - \int_{ - \infty }^t {\bf{F}} (t')dt'$ as the vector potential of the laser field. In fact, the electron wave packet with initial wave vector $\bf{q}$ transforms to a (trajectory-guided) instantaneous crystallographic vector,${\bf{q}} \mapsto {\bf{k}}(t)$, where its  time-dependent trajectory is governed by the laser's vector potential. 
We employ the Liouville von Neumann equation and propagate the reduced density matrix elements ${\rho ^{ij}}$ ($i$,$j$ band indices) to describe our two-band open quantum system in the presence of the relaxation process. 

\begin{equation}
	\label{Eq:Rho_Dot}
	\begin{array}{*{20}{l}}
		{{\partial _t}\tilde \rho _{{\bf{k}}(t)}^{cv} = \frac{{ie}}{\hbar }{\bf{F}}(t) \cdot {\bf{Q}}_{{\bf{k}}(t)}^{cv}\left[ {\tilde \rho _{{\bf{k}}(t)}^{vv} - \tilde \rho _{{\bf{k}}(t)}^{cc}} \right]}\\
		{{\partial _t}\tilde \rho _{{\bf{k}}(t)}^{cc} = 2{\rm{Re}}\left[ {\frac{{ie}}{\hbar }{\bf{F}}(t) \cdot {\bf{Q}}_{{\bf{k}}(t)}^{cv*}\tilde \rho _{{\bf{k}}(t)}^{cv}} \right]}
	\end{array}
\end{equation}
${\tilde \rho ^{ij}} = {\rho ^{ij}}{e^{ - \gamma t}}$ are the unitary transformed elements of the density matrix operator ${\rho ^{ij}}$ with the relaxation rate $\gamma$. Phenomenologically, the relaxation rate has an inverse proportion of the scattering time, ${\gamma _{({\rm{PHz}})}} = \frac{1}{{{T_{({\rm{fs}})}}}}$.  In Eq. \ref{Eq:Rho_Dot}
\begin{equation}
	\label{Eq:Q_cv}
	{\bf{Q}}_{{\bf{k}}(t)}^{cv} = {\mathbfcal A}_{{\bf{k}}(t)}^{cv}{e^{\frac{i}{\hbar }\int_{ - \infty }^t {\left( {E_c^T[{\bf{k}}(t')] - E_v^T[{\bf{k}}(t')]} \right)dt'} }}
\end{equation}
determines the  matrix  element  of  interband interaction where ${\mathbfcal A}_{{\bf{k}}(t)}^{cv}$ is the non-diagonal element of the non-abelian Berry connection, and $E_n^T[{\bf{k}}(t)] = {E_{\rm{n}}}[{\bf{k}}(t)] + e{\bf{F}}(t) \cdot{\mathbfcal{A}^{({\rm{nn}})}}[{\bf{k}}(t)]$ is the ``generalized" energy of ${n^{{\rm{th}}}}$-band, with  $ {{\mathbfcal A}^{(nn)}}$ as the diagonal matrix elements of Berry connection. The mathematical description and analytical expressions for the tensorial components of the Berry connection are presented in Supplementary Information [\hyperlink{SMBerryCon}{S2}].
The generalized band energy incorporates the Bloch eigenenergies [First term], and the topological band dispersion induced by laser waveform [second term].
In other words, it accounts for the dynamic phase [first term], as well as the topological (Berry) phase [second term]. The latter term is critical to characterizing and manipulating the nontrivial phase of condensed matter systems, including the peculiarities observed in the anomalous quantum Hall effect, quantum spin Hall effect, Valley polarization, and High-order harmonic generation.

The rate equations (\ref{Eq:Rho_Dot}) determine the laser-induced electron dynamics;  solving these coupled integro-differential equations, we obtain reciprocal space distribution of electrons in the conduction and valence bands. In Supplementary Information [\hyperlink{SMGauge}{S3}] we set out the correspondence between length-gauge and velocity-gauge in the determination of such quantum electron dynamics. 
Respectively, the generated photocurrent, carrier transfer and other observable are calculated from the density matrix operator. 
The incident optical pulse causes polarization of the system by exciting electrons in the conduction band (CB). Subsequently, a time-dependent electronic current ${\bf{J}}(t) = \left\{ {{J_x}(t),{J_y}(t)} \right\}$ induces in the system.  
Both intraband and interband currents contribute to the total current, and in the density matrix formalism are calculated by the following expressions: 
\begin{equation}
	\label{Eq:current}
	\begin{array}{*{20}{l}}
		{{{\bf{J}}_{{\rm{int}}ra}}(t) = \sum\limits_{n  = c.v} {\sum\limits_{{\rm{BZ}}} {{{\mathbfcal{V}}_n}\left[ {{\bf{k}}(t)} \right]\rho _{{\bf{k}}(t)}^{nn}} } }\\
		{{{\bf{J}}_{{\rm{int}}er}}(t) = 2\sum\limits_{{\rm{BZ}}} {{\mathop{\rm Re}\nolimits} \left( {{\mathbfcal{V}}_{nn'}^*\left[ {{\bf{k}}(t)} \right]\rho _{{\bf{k}}(t)}^{nn}} \right)} }
	\end{array}
\end{equation}
${{\mathbfcal{V}}_n}$ and ${{\mathbfcal{V}}_{nn'}}$, ($n$ and $n'$ interchange between $c$ and $v$) are the matrix elements of the intraband and interband velocity operator, respectively:
\begin{equation}
	\label{Eq:velocity_inter_intra}
	\begin{array}{*{20}{l}}
		{{{\mathbfcal{V}}_n} = \frac{1}{\hbar }{\grad _{\bf{k}}}E_n^T[{\bf{k}}(t)]}\\
		{{{\mathbfcal{V}}_{nn'}}({\bf{k}}) = \frac{i}{\hbar }{\bf{Q}}_{{\bf{k}}(t)}^{nn'}\left[ {E_n^T({\bf{k}}(t)) - E_{n'}^T({\bf{k}}(t))} \right]}
	\end{array}
\end{equation}
${\bf{Q}}_{{\bf{k}}({\bf{q}},t)}^{nn'}$ in Eq. \ref{Eq:velocity_inter_intra} is obtained from Eq. \ref{Eq:Q_cv} and is related to the interband Berry connection.
In fact, intraband velocity contains two terms:  ${\mathbfcal{V}}_n^{{\rm{group}}}({\bf{k}}) = {\grad _{\bf{k}}}{E_n}({\bf{k}})$ is the group velocity with $E_n({\bf k})$ the energy dispersion of respective band, and
${\mathbfcal{V}}_n^{{\rm{anom}}}({\bf{k}}) = e{\grad _{\bf{k}}}\left( {{\bf{F}}(t) \cdot {{\mathbfcal{A}}^{(nn)}}[{\bf{k}}(t)]} \right)$ is the anomalous Hall velocity.  We note that in our formalism, the ``generalized" band energy ${E_n^T}$, carries the topological information of the interacting system. Hence, according to Eq. \ref{Eq:current} and \ref{Eq:velocity_inter_intra}, the intraband velocity (and successively intraband current), and interband velocity (and intraband current) are modulated by the effective band topology. The intra- and inter- band anomalous velocities are plotted in Fig. \ref{Fig_schematic}(c) and (d), respectively. They show the trigonal wrapping, and chiral nature of quasiparticles in the vicinity of the Dirac valleys.

The coherent sum of the intraband and interband currents ${\bf{J}}(t) = {{\bf{J}}^{intra}}(t) + {{\bf{J}}^{inter}}(t)$, consequently results in the HHG as: ${\cal I}(\omega ) = \frac{1}{{\sqrt T }}\int_0^T {J(t){e^{ - i\omega t}}dt} $. Using a discrete Fourier transform for the above equation, we obtain ${\cal I}(\omega ) = \frac{{\Delta t}}{{\sqrt T }}\sum\limits_{n = 1}^N {{J_n}{e^{ - i\omega n\Delta t}}} $ where $T = N\Delta t$ is the total pulse duration. The power spectrum $S(\omega )$, (i.e., harmonic intensity) is computed by ${\cal S}(\omega ) = {\left| {{\cal I}(\omega )} \right|^2} = \frac{{\Delta {t^2}}}{T}{\left| {\sum\limits_{n = 1}^N {{J_n}{e^{ - i\omega n\Delta t}}} } \right|^2}$. The spectral phase of HHG is calculated as ${\cal F}(\omega ) = \arg \left[ {{\cal I}(\omega )} \right] = {\mathop{\rm Im}\nolimits} \left[ {\ln {\cal I}(\omega )} \right]$ 
The time-frequency spectrum is also obtainable via a short-time Fourier transform method (STFT). One such transformation is the Gabor transform: $G(\omega ,t) = \int {d{t^\prime }J({t^\prime }){e^{ - i\omega {t^\prime }}}{e^{ - \frac{{{{(t - {t^\prime })}^2}}}{{2{\tau ^2}}}}}} $ where $\tau $ is the width of the time window. The time-frequency spectrum is then calculated using $S(G(\omega ,t)) = {\left| {G(\omega ,t)} \right|^2}$. 

\section{Results}
\label{sec:result}

To describe the laser-induced process, we employ the following vector potential waveform
\begin{equation}
	\label{Eq:VectorPotential}
	{\bf{A}}(t) = \frac{{{F_0}}}{\omega }{\sin ^2}\left( {\pi t/\tau } \right)\left[ {\frac{1}{{\sqrt {1 + {\varepsilon ^2}} }}\cos (\omega t + {\phi _{{\rm{CEP}}}}){{{\bf{\hat e}}}_x} + \frac{\varepsilon }{{\sqrt {1 + {\varepsilon ^2}} }}\sin (\omega t + {\phi _{{\rm{CEP}}}}){{{\bf{\hat e}}}_y}} \right],
\end{equation}
where ${F_{0}}$ is the peak field amplitude, $\tau $ is the full pulse duration of our sin-squared envelope function, ${{\phi _{{\rm{CEP}}}}}$ the carrier-envelope phase (CEP) and $\varepsilon \in [0,1]$ determines the pulse ellipticity.
We consider a laser pulse of 30-fs duration and the carrier wavelength $\lambda $
is 1.6 $\mu m$, corresponding to $\omega = 0.775$ ${\rm{eV/}}\hbar $. In
all our calculations, we used ${{\phi _{{\rm{CEP}}}}}= 0$.

\begin{figure}
	\begin{center}
		\includegraphics[width=\linewidth]{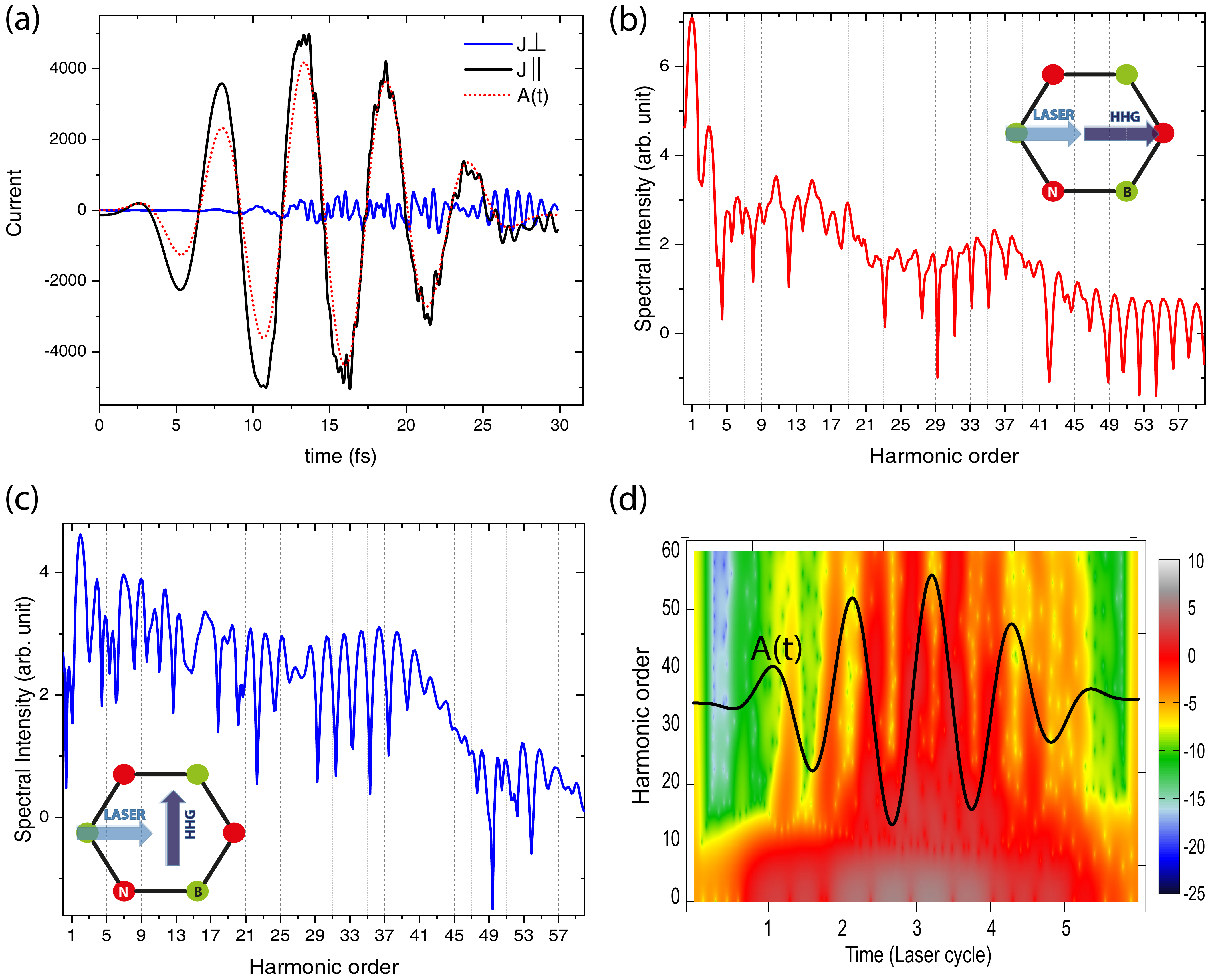}
		\caption{\label{Fig_J_HHG} (a) Time-dependent lightwave current in hBN, parallel and perpendicular to the laser polarization. The dashed line is the field's vector potential. HHG spectra of hBN in the parallel configuration (b), and perpendicular configuration (c). Multiple plateau-cutoff formation is observed. Emission spectra up to 60$^{{\rm{th}}}$ harmonic order (46.5 eV) are plotted from 30 fs linearly polarized driving laser. In part (d), the emission profile for the longitudinal HHG spectra is illustrated. The time profile of the vector potential is also shown. The time window for the Gabor transformation is taken to be 0.25 fs. }
	\end{center}
\end{figure}

\subsection{\label{Sec:Field_amplitude} High-harmonic generation}
HHG is a nonperturbative nonlinear probe of ultrafast charge dynamics induced by the strong-field lasers. As we demonstrate in this section, HHG is a sensitive tool for probing nontrivial phases in topological materials.  
We look into the laser-induced current and higher-order harmonic spectra of hBN as a prototypical candidate of 2D systems with intrinsic spatial inversion symmetry breaking. Both longitudinal and transverse nonlinear responses are taken into account.
Fig. \ref{Fig_J_HHG} (a) plots the generated current density parallel to the laser polarization (${J_\parallel }$) as well as perpendicular configuration (${J_ \bot }$). The residual current is originated from breaking the the system's spatiotemporal symmetry by the few-cycle nature of the laser.
In Fig. \ref{Fig_J_HHG} (b) and (c), the harmonic intensity is plotted respectively for longitudinal and transverse spectra. 
Due to the presence of mirror plane \cite{Heinz_Reis2016}, even harmonics are absent in the parallel excitation. In the perpendicular direction, however, we observe even Harmonics, which is originated from the quantum Hall current induced by the electron's anomalous velocity.
The multiple plateau-cutoff structures, as a characteristic of solid-state HHG, are also observed.
The time-frequency profile of the HHG spectra for longitudinal response is density plotted in Fig. \ref{Fig_J_HHG}(d) by taking Gabor transformation of the time-dependent current. The chirps of attosecond emission are observed, which are created at the extrema of the vector potential. 
The sign and magnitude of the attosecond chirp can be controlled by the laser properties such as wavelength, amplitude, and carrier-envelope phase.
As a side remark, it is possible to generate an ultrabright, and intense single-cycle attosecond pulses (SAP) in this system with a duration shorter than 200 attoseconds. SAP has been an actively researched in atoms and noble gases using polarization gating and Double optical gating \cite{Lin_Attosecond_2018}. The tendency is currently shifted toward solids.

\subsection{\label{Sec:relaxation} Lightwave anomalous quantum Hall effect } 

In a direct bandgap crystal with broken inversion symmetry, the two degenerate valleys can be distinguished by pseudovector identities; namely, the Berry connection and Berry curvature, which must take opposite values at the time-reversal pair of Dirac valleys.
The valley contrasted Berry curvatures can couple to external electric fields, giving rise to the Quantum Valley Hall effect (QVHE). It greatly influences key properties of electrons in solids, including electric polarization, anomalous Hall conductivity, and the nature of the topological insulating state.

\begin{figure}
	\begin{center}
		\includegraphics[width=\linewidth]{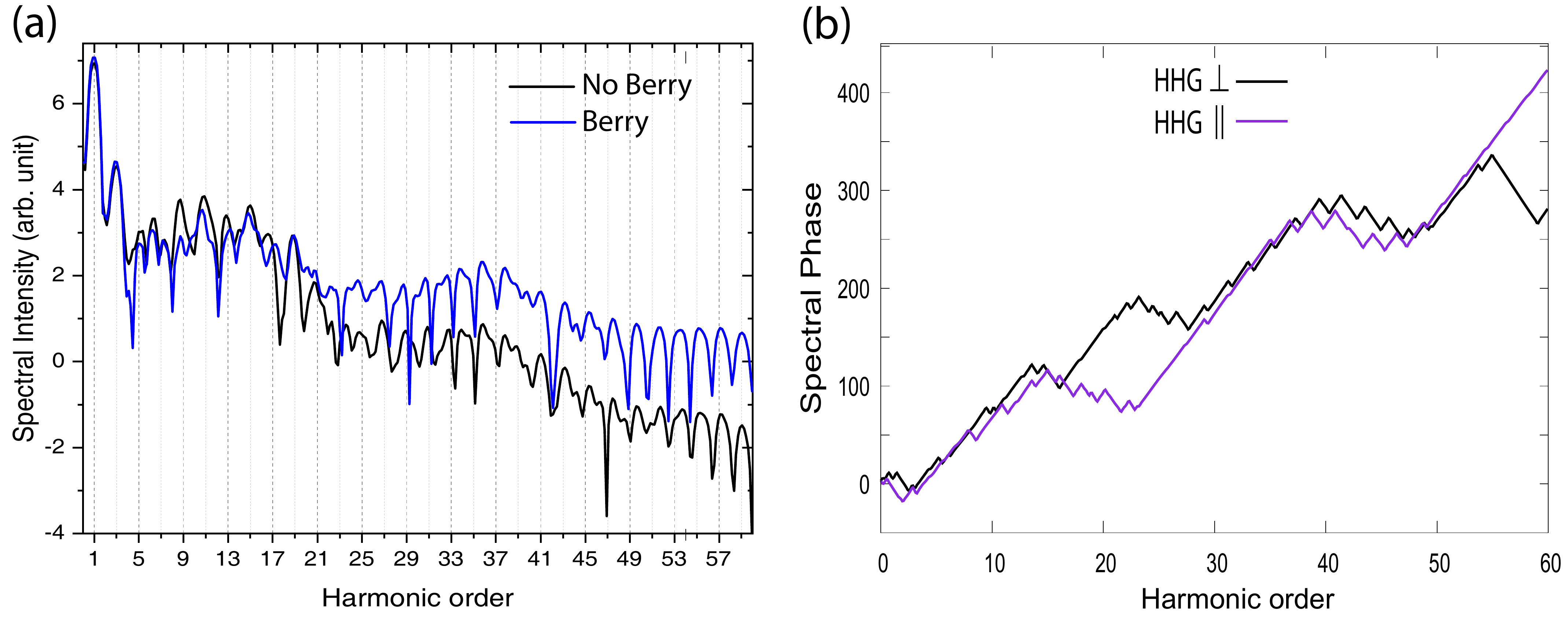}
		\caption{\label{Fig_Berry} (a) Depicts the effect of Berry phase in HHG intensity; the (longitudinal) HHG is compared with the case where we intentionally disabled the contribution of anomalous quantum Hall effects originated from the Berry connection and curvature.  (b) The Phase of HHG spectra, ${\cal F}(\omega )$, for longitudinal and transverse configurations. The multiple plateau-cutoff structures of the emitted harmonics is apparent from sharp phase jumps at the harmonic orders associated with the cutoff energies.  }
	\end{center}
\end{figure}

Fig. \ref{Fig_Berry} (a) demonstrates the role of Bloch band topology in HHG yield; the black line plots emission intensity where we deliberately switched off the contribution of Berry connection term is in Eqs. \ref{Eq:current}-\ref{Eq:velocity_inter_intra}. The harmonic yield drops considerably in the plateau and cutoff regions where the intraband current is dominant.
Furthermore, the (multiple) plateaux and cutoff structure of the HHG can be evidently seen as abrupt phase jump across the HHG spectral phase, ${\cal F}(\omega )$, at harmonic orders associated with cutoff energies [Fig. \ref{Fig_Berry} (b)].

\begin{figure}
	\begin{center}
		\includegraphics[width=\linewidth]{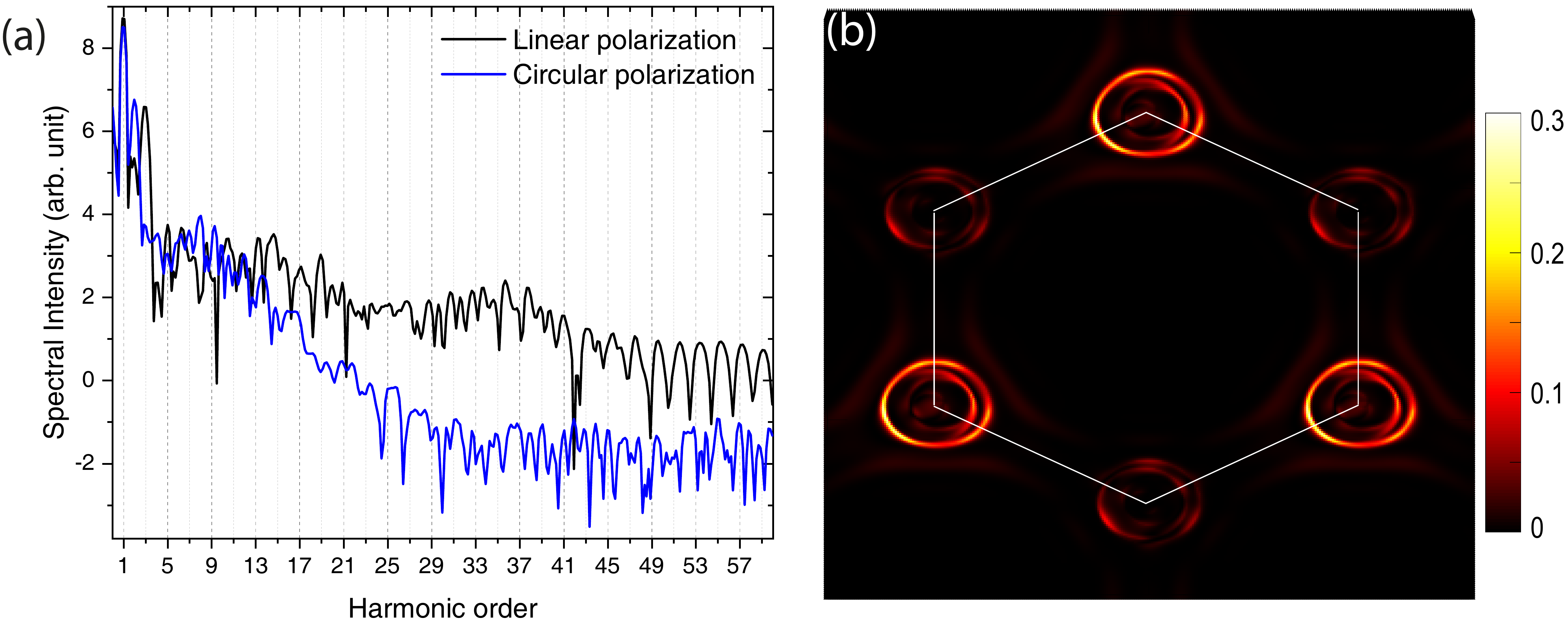}
		\caption{\label{Fig_Cir} (a) HHG spectra for hBN driven by circularly-polarized field are compared with the emitted HHG from linear pulse (b) Momentum- resolved residual distribution of electrons in the conduction bands for the circularly polarized pulse after the light-matter interaction. The sin-squared pulse has a field amplitude of 0.6 ${\rm{V/\AA}}$. The valley-contrasting excitation at the corresponding Dirac cones is associated with nonzero Berry curvature and an intrinsic pseudo-magnetic moment near the Dirac cones.   }
	\end{center}
\end{figure}

We have studied, so far, the HHG mechanism in broken inversion symmetry systems. 
At this point, we turn into the time-reversal symmetry (TRS) broken systems. For that reason, we exploit the polarization dependence of HHG to examine.  Fig.\ref{Fig_Cir} (a) compares the circularly polarized pulse HHG with respect to the harmonic spectra of linear polarization. Reciprocal space population distribution of electrons in CB for a circular pulse is shown in Fig. \ref{Fig_Cir} (b). The peak field amplitude is $F_0=0.6 {\rm{V/\AA}}$.
A chiral pulse breaks TRS; subsequently, in hBN - or equivalently any massive 2D Dirac materials- one valley acquires a high CB population while the other valley has almost zero CB population. Such broken symmetries driven by the chiral optical field is also signified in the HHG spectra through the emergence of even harmonics [see Fig. \ref{Fig_Cir} (a)]. 
Such a structure is absent in pristine graphene \cite{Kelardeh_circular} and is attributed to the $\pm \pi $- Berry phase, respectively, at the ${\bf{K}}$ and ${\bf{K'}}$ valleys. However, in the broken inversion symmetry systems, the Berry phase is modular and gap-dependent. The mathematical representation of this concept is introduced in Supplementary Information [\hyperlink{SMSpinor}{S1}] by taking an analogy with spinors on (parametric) Bloch sphere.

\begin{figure}
	\begin{center}
		\includegraphics[width=\linewidth]{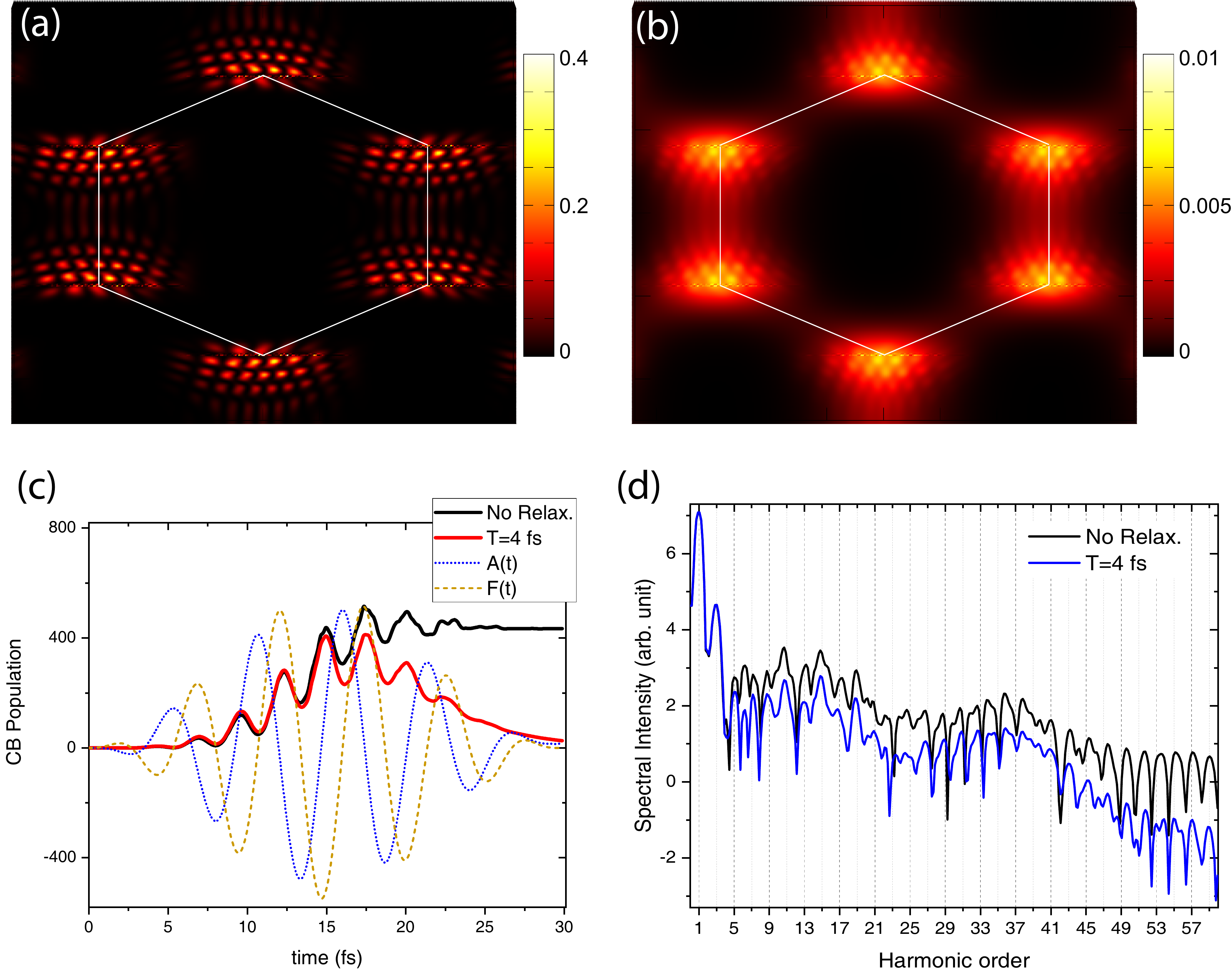}
		\caption{\label{Fig_Relax} (a) conduction band (CB) population distribution (${\rho _{{\bf{k}}(t \to {t_f})}^{cc}}$ ) for non-scattering simulation.  The laser is linearly polarized at the plane of hBN  with $F_0=0.6 {\rm{V/\AA}}$ and ${\varphi _{{\rm{CEP}}}} = 0$. The borderline for the first Brillouin zone (BZ) is shown. Residual CB excitation in the extended BZ for an alternative scenario with dephasing time $T=4$ fs. (c) Time-dependent total population is plotted within the incident field for non-scattering electron dynamics versus the $T=4$ fs dephasing time. The laser field and vector potential are also plotted with dashed lines. The kick-like excitation is due to the interplay between interband dipole transition and intraband Bloch oscillation and is denoted as the carrier-wave Rabi flopping. (d) the HHG spectra for non-scattering and $T=4$ fs dephasing time are contrasted.  }
	\end{center}
\end{figure}

\subsection{\label{Sec:relaxation} Many-particle effects on HHG} 

Now, we turn into the impact of the relaxation process on momentum-resolved excitation dynamics of electrons and the HHG yield. When we disregard the relaxation rate in Eq. \ref{Eq:Rho_Dot}, the momentum resolved CB population demonstrates the interference fringes along the electron trajectory and distributes asymmetrically in the ${\bf{k}}$-space [see Fig. \ref{Fig_Relax} (a)]. Alternatively, the residual CB population under the influence of a fast dephasing time, $T=4$ fs, is plotted in Fig. \ref{Fig_Relax} (b). The laser is linearly polarized with sin-squared waveform, field amplitude $F_0=0.6 {\rm{V/\AA}}$, and ${\phi _{{\rm{CEP}}}} = 0$.
The short-time electron scattering affects the magnitude of the electron wave packet as well as its phase; it smears the excitation distribution along the electron trajectory. As a result, the excitation relaxes to the ground states and the total CB population $\sum\limits_{{\rm{BZ}}} {\rho _{{\bf{k}}(t)}^{cc}} $ is nearly zero at the end of the pulse [ see Fig. \ref{Fig_Relax} (c)]. The oscillatory behavior of electron distribution recalls carrier-wave Rabi flopping (CWRF). CWRF takes place in the intermediate Keldysh regime \cite{Yakovlev_Colloquium_2018} where interband transition (Rabi oscillation) is strongly coupled to intraband motion. In our Dirac system, such a process mainly occurs when electrons pass through or near the Dirac point where the transition dipole moment is strong and electron wave packet modulates very rapidly.  

In Fig. \ref{Fig_Relax} (d) we compared the HHG spectra for fast dephasing time ($T=4$ fs) with respect to the non-scattering case for linear polarization. 
Compared to the non-scattering case, the fast dephasing scenario has roughly one order of magnitude lower HHG yield up until the third plateau (47$\omega $). Within this range, the profile of HHG for both cases are nearly the same. In the third plateau (harmonic order $ > 47\omega $), however, for $T=4$ the signal is quite noisy and indistinct contrary to the non-dephasing calculation.

\section{Conclusion}

This paper sets forth a gauge-invariant theory for optically-drive electrons in two-dimensional (2D) graphene-like systems on the onset of generalized density matrix formalism. 
Our presented model is based on the Liouville von Neumann equation, and revises the semiclassical transport model that overlooks the topology of the electron wave function.
The topology of Bloch eigenstates governed by Berry's trilogy $-$ connection, curvature and phase $-$, shapes the directionality and the attosecond timing of electron injection into the conduction band and photogenerated currents.
We self-consistently incorporate the band topology and topological phase transitions in the nonperturbative optical response of the strong-field light-matter interaction. 

Namely, we have shown that HHG is a sensitive observable to probe different topological phases and phase transitions in topological materials. The HHG emission spectra represent an attosecond chirp dissimilar to the atomic and molecular HHG.
The multiple plateau-cutoff structures are observed in both parallel and perpendicular directions to the non-chiral incident field polarization. The transverse photoemission spectra form even harmonics owing to the anomalous quantum Hall effect induced by the electron's anomalous velocity. 
Likewise, the spectral phase of the HHG represents susceptibility to the band structure through the stepping phase jumps at the harmonic orders associated with the cutoff energies.

The chiral field, on the other hand, induces an extremely robust valley polarization in 2D solids owing to the Bloch band topology. Such a symmetry breaking is also characterized in the HHG profile of hBN. Notably, we assert and mathematically validate that the photoinduced valley polarization fundamentally occurs in all 2D graphene-like materials irrespective of their chemical composition, spin-orbit coupling strength, and so on.
They enable topologically-robust valleytronic devices to write-in and read-out signals with the petahertz data rate.
The result and theory presented in this letter stimulate developments in the coherent control of solids, as well as topological strong-field optics of semimetals, insulators, and superconductors.

\hypertarget{SMSpinor}{\section*{S1: Universality manifestation of 2D systems with broken inversion symmetry}}
\label{SM_Spinor}

A direct implication of Berry’s phase in 2D Dirac systems can be understood in the context of spin-1/2 spinors. Here we mathematically prove that the bandgap (i.e., effective mass) by breaking the inversion symmetry shift the Berry phase to values different from $\pi$. This is critical to capture the signatures of Berry phase in the momentum resolved excitation of electron wave packet.
In the near Dirac approximation, the generic Hamiltonian of 2-band system reads:
\begin{equation}
	{H_{2D}} = {k_x}{\sigma _x} + {k_y}{\sigma _y} + m{\sigma _z} = \left( {\begin{array}{*{20}{c}}
			m&{\gamma \left| k \right|{e^{ - i{\varphi _k}}}}\\
			{\gamma \left| k \right|{e^{i{\varphi _k}}}}&{ - m}
	\end{array}} \right)
\end{equation}
if $m = 0$ we get the eigenenergies and eigenstates of gapless 2D semimetal systems (such as intrinsic graphene):
\begin{equation}
	\label{eq:Graphene_ND}
	{{\rm{E}}_{c,v}} =  \pm \gamma \left| {\bf{k}} \right|,{\mkern 1mu} {\mkern 1mu} {\mkern 1mu} {\mkern 1mu} {\Psi _{c,v}} = \frac{1}{{\sqrt 2 }}\left( {\begin{array}{*{20}{c}}
			{ \pm 1}\\
			{{e^{i{\varphi _{\bf{k}}}}}}
	\end{array}} \right),
\end{equation}
$+$ and $-$ signs denote the states above (CB) and below (VB) the Dirac point, respectively. The two-component vector in Eq.\ref{eq:Graphene_ND} can be viewed as a result of the the spin-1/2 rotation operator. In spherical coordinate, the Hamiltonian of a spin-$1/2$ particle in an arbitrary direction ${\bf{\hat n}}({\bf{k}}) = \left( {\sin \theta \cos \varphi ,\sin \theta \sin \varphi ,\cos \theta } \right)$ is
\begin{equation}
	{H_{spinor}} = \varepsilon {\bf{\sigma }} \cdot {\bf{\hat n}} = \left( {\begin{array}{*{20}{c}}
			{\cos \theta }&{\sin \theta {e^{ - i\varphi }}}\\
			{\sin \theta {e^{i\varphi }}}&{ - \cos \theta }
	\end{array}} \right)
\end{equation}
where ${\sigma ^i}$ are the Pauli matrices. The eigenvalues and eigenstates read:
\begin{equation}
	\label{Eq:spinor}
	{{\rm{E}}_{ \uparrow , \downarrow }} =  \pm \varepsilon ,{\mkern 1mu} {\mkern 1mu} {\mkern 1mu} {\mkern 1mu} {\psi _{ \uparrow , \downarrow }} = \left( {\begin{array}{*{20}{c}}
			{ \pm \cos (\theta /2)}\\
			{\sin (\theta /2){e^{i\varphi }}}
	\end{array}} \right),
\end{equation}
Comparing Eq. \ref{eq:Graphene_ND} and Eq. \ref{Eq:spinor}, one can infer that Eq. \ref{eq:Graphene_ND} is a special case of the spin-$1/2$ problem with $\theta=\pi /2$.
In fact, introducing a bandgap is analogous to giving a finite $\theta$ that rotates wave function on the Bloch sphere. Respectively, one obtains explicit expressions for Berry connection, Berry curvature, and Berry phase as follows: 
\begin{itemize}
	\item \textbf{Berry Connection:} ${{{\cal A}_{ \uparrow  \uparrow }} = i\langle {\psi _ \uparrow }|\grad \left| {{\psi _ \uparrow }} \right\rangle  = {A_\varphi }\hat \varphi  =  - \frac{{\tan (\theta /2)}}{{2k}}\hat \varphi }$
\end{itemize}  

\begin{itemize}
	\item \textbf{Berry curvature:} ${\Omega  = {\rm{ }}\grad  \times {{\cal A}_{ \uparrow  \uparrow }} = \frac{{{\bf{\hat k}}}}{{k\sin \theta }}{\partial _\theta }\left( {k\sin \theta {A_\varphi }} \right) =  - \frac{{{\bf{\hat k}}}}{{2{k^2}}}}$
\end{itemize}

\begin{itemize}
	\item \textbf{Berry Phase:} $\gamma  = \oint {{{\cal A}_{ \uparrow  \uparrow }} \cdot d{\bf{k}} = \oint {{A_\varphi }d\varphi (k\sin \theta )} }  = \pi \left( {1 - \cos \theta } \right)$
\end{itemize}


This indicates that for pristine graphene which preserve inversion symmetry, $\gamma  =  \pm \pi $ ($ \pm $ corresponding to the ${\bf{K}}$ and ${\bf{K'}}$ valleys). 
In other words, the $2\pi $ rotation of pseudospin brings about a phase factor of ${e^{i\pi }}$ in the electronic wave function. Such a phase factor manifests itself as a phase jump in the momentum distribution of the electron wave function when an electron makes a cyclic trajectory in the parametric space by the circular electric field  ( see Ref. \cite{Kelardeh_circular,Kelardeh_SPIE_Circular} ). 
However, the CB population distribution, as an observable, is insensitive to such a phase transformation (${e^{2i\pi }} = 1$).   
On the other hand, for a broken inversion symmetry system, such as hBN, topological crystalline insulators, gapped graphene, TMDs, the Berry phase differs from $\pi$ and allows us to detect the signature of phase shift directly in the population distribution. Such a phase jump has been previously detected by the author in the context of graphene superlattices \cite{Kelardeh_superlattice,Kelardeh_SPIE_Superlattice}, and equivalently can be measured in Moiré heterostructures \cite{Caldwell_2019} or twisted bilayer graphene \cite{Jarillo_Herrero_twisted_BG}.

The association of energy bandgap with Bloch sphere, postulates a critical conception in the condensed matter physics: \textit{ the effective bandgap universally governs Valleytronics in 2D crystals and factors such as spin-orbit coupling, and chemical composition and etc., have fewer impacts in this regard.  }

\hypertarget{SMBerryCon}{\section*{S2: Matrix elements of the non-Abelian Berry connection}}
\label{SM_BerryConnection}



Let's initially solve the eigenenergies and eigenstates of a generic 2D hexagonal crystal with bandgap $m$ in tight-binding model:
\begin{equation}
	\label{GG_Hamiltonian}
	{{\cal{H}}_0} = \left( {\begin{array}{*{20}{c}}
			m&{g({\bf{q}})}\\
			{{g^*}({\bf{q}})}&{ - m}
	\end{array}} \right)
\end{equation}
where $g({\bf{q}}) = \gamma \left[ {\exp \left( {i\frac{{a{q_x}}}{{\sqrt 3 }}} \right) + 2\exp \left( { - i\frac{{a{q_x}}}{{2\sqrt 3 }}} \right)\cos \left( {\frac{{a{q_y}}}{2}} \right)} \right]$, with $\gamma$ as the hopping potential, and $a$ the lattice constant.
We define $g({\bf{q}}) = \left| {g({\bf{q}})} \right|{e^{i{\varphi _{\bf{q}}}}}$, with ${\varphi _{\bf{q}}} = \arg \left[ {g({\bf{q}})} \right]$. Later we drop the ${\bf{q}}$-indices to better read. 
The eigenvalues and eigenstates of Eq. \ref{GG_Hamiltonian} can be calculated as
\begin{equation}
	\label{Eq:Bloch_eigensystem}
	\begin{array}{l}
		{E_ \pm } =  \pm \sqrt {{m^2} + {{\left| g \right|}^2}} \\
		\left| {{\phi ^ \pm }} \right\rangle  = \frac{1}{{\sqrt {1 + \Delta _ \pm ^2} }}\left( {\begin{array}{*{20}{c}}
				{{\Delta _ \pm }}\\
				{{e^{ - i\varphi }}}
		\end{array}} \right)
	\end{array}
\end{equation}
where ${\Delta _ \pm } = \frac{1}{{\left| g \right|}}\left( {m \pm \sqrt {{m^2} + {{\left| g \right|}^2}} } \right)$. $\rm{ + , - }$ respectively correspond to the conduction ($ + $), and valence ($ - $) bands. 
In general, the Berry connection is a non-abilian tensor with tonsorial elements as
\begin{equation}
	\label{Eq:Berry_elements}
	{{\mathbfcal A}^{nm}} = i\left\langle {{\phi ^n}|{\grad _{\bf{q}}}|{\phi ^m}} \right\rangle 
\end{equation}
in which $n,m $ toggle between $c,v$. The diagonal terms are intraband, and the off-diagonals are the intraband Berry connections. The latter determines the matrix elements of transition dipole operator (TDO) \cite{Kelardeh_Wannier} as ${\bf{D}}_{{\bf{k}}(t)}^{cv} = e{\mathbfcal A}_{{\bf{k}}(t)}^{cv}$. TDO determines optical selection rules between the conduction and valence bands at the instantaneous crystal momentum ${\bf{k}}(t) = {\bf{q}} - {\bf{A}}(t)$. 
Substituting the Bloch eigenstates (Eq. \ref{Eq:Bloch_eigensystem}) into Eq. \ref{Eq:Berry_elements}, an analytical expression for the matrix elements of the intraband and interband Berry connection is obtained

\begin{equation}
	\label{Eq:BerryConnection}
	\begin{array}{l}
		{{\mathbfcal A}^{(cc),(vv)}} = \frac{{{{\left| g \right|}^2}\grad \varphi }}{{{u_{c,v}}}}\\
		{{\mathbfcal A}^{cv}} = \frac{{\left| g \right|\grad \varphi }}{{2{E_c}}} - i\frac{{m\grad \left| g \right|}}{{2E_c^2}}
	\end{array}
\end{equation}
with ${u_{c,v }} = 2{E_{c,v }}\left( {m + {E_{c,v }}} \right)$. Taking the explicit expressions for $\grad{\varphi}$ and $\grad{\left| {g} \right|} $ the $x$ and $y$ components of the Berry connections are derived on the onset of tight-binding model: 

\begin{equation}
	\label{Eq:BerryCon_TB}
	\begin{array}{*{20}{l}}
		{{\cal A}_x^{cc,vv} = \frac{{a{\gamma ^2}}}{{\sqrt 3 }}\frac{{1 + {c_0}({c_3} - 2{c_0})}}{{{u_{c,v}}}},\,\,{\cal A}_y^{cc,vv} = a{\gamma ^2}\frac{{{s_0}{s_3}}}{{{u_{c,v}}}}}\\
		{{\cal A}_x^{cv} = \frac{{a{\gamma ^2}}}{{2\sqrt 3 {E_c}\left| g \right|}}\left[ {1 + {c_0}({c_3} - 2{c_0})} \right] + i\frac{{\sqrt 3 m a{\gamma ^2}}}{{2E_c^2\left| g \right|}}{c_0}{s_3},\,\,{\cal A}_y^{cv} = \frac{{a{\gamma ^2}}}{{2{E_c}\left| g \right|}}{s_0}{s_3} + i\frac{{m a{\gamma ^2}}}{{2E_c^2\left| g \right|}}{s_0}\left( {{c_3} + 2{c_0}} \right)}
	\end{array}
\end{equation}

where ${c_0} = \cos \left( {a{k_y}/2} \right)$, ${s_0} = \sin \left( {a{k_y}/2} \right)$, ${c_3} = \cos \left( {\sqrt 3 a{k_x}/2} \right)$,  ${s_3} = \sin \left( {\sqrt 3 a{k_x}/2} \right)$. 

\hypertarget{SMGauge}{\section*{S3: Gauge invariance in the laser-induced system}}
\label{SM_Length_Velocity_Gauge}

If we assume that in ultrafast regime, the scattering processes do not develop on sub-cycle time scales and thereby negligible, the solution of optically-induced electron dynamics is given by the following single-particle eigenvalue problem:
\[i\hbar \frac{d}{{dt}}\left| \Psi  \right\rangle  = \left[ {\frac{{{{\left( {{\bf{p}} - {\bf{A}}({\bf{r}},t)} \right)}^2}}}{{2m}} + e\varphi ({\bf{r}},t) + V({\bf{r}})} \right]\left| \Psi  \right\rangle \]
infinite combinations  of $\bf{A}$  and  $\varphi$ will give rise to the solution of the Hamiltonian. In particular, one can identify two  independent choices: 

the vector-potential gauge (\underline{velocity gauge}):
\[A({\bf{r}},t)\mathop  \to \limits^{{\rm{dipole}}} A(t) =  - \int_{ - \infty }^t {F(t')dt'} ,\,\,\,\,\,\,\,\varphi ({\bf{r}},t) = 0\]

and the scalar-potential gauge (\underline{Length gauge}):
\[{A_L}({\bf{r}},t) = 0,\,\,\,\,\,\,\,{\varphi _L}({\bf{r}},t) =  - {\bf{F}}(t) \cdot {\bf{r}}\]
the connection between the two representation is obtained by the following gauge transformation:
\[{{\bf{A}}_L} = {\bf{A}} + \grad \chi ,\,\,\,\,{\varphi _L} = \varphi  - \frac{{\partial \chi }}{{\partial t}}\]
with $\chi  =  - {\bf{A(t)}} \cdot {\bf{r}}$.  Thus if $\Psi $ is a solution of 
\[i\hbar {{\dot \Psi }_V} = \left\{ {\frac{{{{\left( {{\bf{p}} - {\bf{A}}({\bf{r}},t)} \right)}^2}}}{{2m}} + V({\bf{r}})} \right\}{\Psi _V}\]
with ${\bf{A}} =  - \int_{ - \infty }^t {{\bf{F}}(t')dt'} $,
the corresponding solution of
\[i\hbar {{\dot \Psi }_L} = \left\{ {\frac{{{{\bf{p}}^2}}}{{2m}} + V({\bf{r}}) - e{\bf{F}}(t) \cdot {\bf{r}}} \right\}{\Psi _L}\]  
is ${\Psi _L} = \Psi_V {e^{ - i{\bf{A}} \cdot {\bf{r}}}}$.

\vspace{0.5cm}
\begin{itemize}
	\item \textbf{Length Gauge:} 
\end{itemize}
Let us consider the solution to the dynamical equation for a two-band crystal in the length gauge in the presence of an intense optical field. 
the Hamiltonian operator ${\cal H}$ given by $ 	{{\cal H} = {\cal H}_0 + e{\bf{F}}(t) \cdot {\bf{r}}} $.
Taking the Bloch eigenfunctions $\phi _{\bf{k}}^{(c,v)}({\bf{r}}) = {e^{i{\bf{k}} \cdot {\bf{r}}}}{\hat u_{c,v}}({\bf{k}})$ of field-free Hamiltonian ${\cal H}_0$ as the basis, we express the general solution in the form $	\Psi_{\mathbf{L}} (\mathbf{r},t) = \sum_{\mathbf{k}} \left[\alpha_v(\mathbf{k}) \phi^{(v)}_{\mathbf{k}} (\mathbf{r}) +
\alpha_c(\mathbf{k}) \phi^{(c)}_{\mathbf{k}} (\mathbf{r})      \right]~.
\label{PsiVC}$ and obtain
\begin{equation}
	\label{Eq:temp1}
	i\hbar \frac{d}{{dt}}\sum\limits_{{\bf{k'}}} {\left[ {{\alpha _v}({\bf{k'}})\phi _{{\bf{k'}}}^{(v)}({\bf{r}}) + {\alpha _c}({\bf{k'}})\phi _{{\bf{k'}}}^{(c)}({\bf{r}})} \right]}  = \left( {{H_0} + e{\bf{F}} \cdot {\bf{r}}} \right)\sum\limits_{{\bf{k'}}} {\left[ {{\alpha _v}({\bf{k'}})\phi _{{\bf{k'}}}^{(v)}({\bf{r}}) + {\alpha _c}({\bf{k'}})\phi _{{\bf{k'}}}^{(c)}({\bf{r}})} \right]} 
\end{equation}
multiplying both sides of Eq. \ref{Eq:temp1} by  $\phi^{(v)*}_{\mathbf{k}} (\mathbf{r})$ and then integrate it by $ \mathbf{r}$ to get
\begin{equation}
	\label{Eq:temp2}
	i\hbar \frac{{d{\alpha _v}({\bf{k}})}}{{dt}} = {E_v}({\bf{k}}){\alpha _v}({\bf{k}}) + e\sum\limits_{{\bf{k'}}} {{\alpha _v}} ({\bf{k'}})\int d {\bf{r}}\phi _{\bf{k}}^{(v)*}({\bf{r}})({\bf{F}} \cdot {\bf{r}})\phi _{{\bf{k'}}}^{(v)}({\bf{r}}) + e\sum\limits_{{\bf{k'}}} {{\alpha _c}} ({\bf{k'}})\int d {\bf{r}}\phi _{\bf{k}}^{(v)*}({\bf{r}})({\bf{F}} \cdot {\bf{r}})\phi _{{\bf{k'}}}^{(c)}({\bf{r}})
\end{equation}
where we have taken the orthogonality condition of the Bloch bands.
note that the orthogonality and completeness of the Bloch eigenfunctions read:
\[\begin{array}{l}
	\int {d{\bf{r}}\phi _{\bf{k}}^{(n)*}({\bf{r}})\phi _{{\bf{k'}}}^{(m)}({\bf{r}})}  = {\delta _{nm}}\delta ({\bf{k}} - {\bf{k'}})\\
	\sum\limits_n {\int {d{\bf{k}}} } \phi _{\bf{k}}^{(n)*}({\bf{r}})\phi _{\bf{k}}^{(n)}({\bf{r'}}) = \delta \left( {{\bf{r}} - {\bf{r'}}} \right)
\end{array}\]
therefore, it is straightforward to get the following identity
\begin{equation}
	\label{delta_kkp}
	\sum\limits_{{\bf{k'}}} {f({\bf{k'}})\int {d{\bf{r}}{e^{i({\bf{k}} - {\bf{k'}}) \cdot {\bf{r}}}}} }  = \sum\limits_{{\bf{k'}}} {f({\bf{k'}})\delta ({\bf{k}} - {\bf{k'}})}  \simeq \int {f({\bf{k'}})\delta ({\bf{k}} - {\bf{k'}})d{\bf{k'}}}  = f({\bf{k}})
\end{equation}
we can further expand the right hand side of Eq. \ref{Eq:temp2} to
\[\begin{array}{l}
	i\hbar \frac{{d{\alpha _v}({\bf{k}})}}{{dt}} = {E_v}({\bf{k}}){\alpha _v}({\bf{k}}) + ie{\bf{F}} \cdot \sum\limits_{{\bf{k'}}} {{\alpha _v}} ({\bf{k'}})\int d {\bf{r}}{e^{i({\bf{k}} - {\bf{k'}}) \cdot {\bf{r}}}}\left[ {u_v^*({\bf{k}}){\grad _{{\bf{k'}}}}{u_v}({\bf{k'}})} \right]\\
	\,\,\,\,\,\,\,\,\,\,\,\,\,\,\,\,\,\,\,\,\,\,\,\,\,\,\,\,\,\,\,\,\,\,\,\,\,\,\,\,\,\,\,\,\,\,\,\,\,\,\,\,\, + ie{\bf{F}} \cdot \sum\limits_{{\bf{k'}}} {{\alpha _c}} ({\bf{k'}})\int d {\bf{r}}{e^{i({\bf{k}} - {\bf{k'}}) \cdot {\bf{r}}}}\left[ {u_v^*({\bf{k}}){\grad _{{\bf{k'}}}}{u_c}({\bf{k'}})} \right]
\end{array}\]
taking Eq. \ref{delta_kkp} we finaly obtain:
\begin{equation}
	\label{Eq:Rate_alphaV}
	i\hbar \frac{{d{\alpha _v}({\bf{k}})}}{{dt}} = \left( {{E_v}({\bf{k}}) + ie{\bf{F}}(t) \cdot {{\bf{A}}_{vv}}({\bf{k}})} \right){\alpha _v}({\bf{k}}) + ie{\bf{F}}(t) \cdot {{\bf{D}}_{vc}}{\alpha _c}({\bf{k}})
\end{equation}
respectively for  ${\alpha _c}({\bf{k}})$ we get
\begin{equation}
	\label{Eq:Rate_alphaC}
	i\hbar \frac{{d{\alpha _c}({\bf{k}})}}{{dt}} = \left( {{E_c}({\bf{k}}) + ie{\bf{F}}(t) \cdot {{\bf{A}}_{cc}}({\bf{k}})} \right){\alpha _c}({\bf{k}}) + ie{\bf{F}}(t) \cdot {{\bf{D}}_{cv}}{\alpha _v}({\bf{k}})
\end{equation}
Now taking the transformation ${\alpha _{c,v}}({\bf{k}}) = {\beta _{c,v}}({\bf{k}}){e^{\frac{1}{{i\hbar }}\int_{ - \infty }^t {E_{c,v}^T({\bf{k}},t')dt'} }}$  with $E_{c,v}^T({\bf{k}},t) = {E_{c,v}}({\bf{k}}) + e{\bf{F}}(t) \cdot {{\bf{A}}_{cc,vv}}({\bf{k}})$ we have
\begin{equation}
	\label{Eq:Rate_houston}
	\left\{ {\begin{array}{*{20}{l}}
			{i\hbar {{\dot \beta }_c} = e{\bf{F}}(t) \cdot {{\bf{Q}}_{cv}}{\beta _v}}\\
			{i\hbar {{\dot \beta }_v} = e{\bf{F}}(t) \cdot {\bf{Q}}_{cv}^*{\beta _c}}
	\end{array}} \right.
\end{equation}
where ${{\bf{Q}}_{cv}} = {e^{\frac{1}{{i\hbar }}\int_{ - \infty }^t {\left( {E_v^T({\bf{k}},t') - E_c^T({\bf{k}},t')} \right)dt'} }}{{\bf{D}}_{cv}}$
In transition from Eq. (\ref{Eq:Rate_alphaV} and \ref{Eq:Rate_alphaC}) to Eq. (\ref{Eq:Rate_houston}) we utilized Leibnitz' formula:
\[{\frac{d}{{dt}}\int_{{\Gamma _1}(t)}^{{\Gamma _2}(t)} {F(x,t)dx = \int_{{\Gamma _1}(t)}^{{\Gamma _2}(t)} {\frac{{\partial F}}{{\partial t}}dx + F({\Gamma _2},t)\frac{{d{\Gamma _2}}}{{dt}} - F({\Gamma _1},t)\frac{{d{\Gamma _1}}}{{dt}}} } }\]

\vspace{0.5cm}
\begin{itemize}
	\item \textbf{Velocity gauge:} 
\end{itemize}
we have
\begin{equation}
	\label{Eq:TDSE_VG}
	i\hbar {{\dot \Psi }_{\bf{k}}}({\bf{r}},t) = \left\{ {\frac{{{{\left( {{\bf{p}} - {\bf{A}}({\bf{r}},t)} \right)}^2}}}{{2m}} + V({\bf{r}})} \right\}{\Psi _{\bf{k}}}({\bf{r}},t)
\end{equation}
similar to the time-evolution operator, $\hat U = {e^{ - i{\cal H}t/\hbar }}$, which displaces the wavefunction in time, we can take the unitary shift operator of crystallographic wave vector (${\bf{k}} \mapsto {\bf{\tilde k}}(t) = {\bf{k}} - {\bf{A}}(t)$), as ${e^{i{\bf{r}} \cdot {\bf{A}}(t)}}\left| {n,{\bf{\tilde k}}(t)} \right\rangle  = \left| {n,{\bf{k}}} \right\rangle $.
Hence
\begin{equation}
	\label{Eq:TDSE_modified}
	i\hbar \dot \Psi  = \frac{{{{{\bf{\tilde p}}}^2}}}{{2m}} + V({\bf{r}}) = \tilde H\Psi 
\end{equation}
For an initial crystal momentum $\bf{k}$, ${{\Psi _{\bf{k}}}({\bf{r}},t)}$ can be expanded as
\begin{equation}
	\label{Eq:Psi_expansion_ket}
	\left| {{\Psi _{\bf{k}}}({\bf{r}},t)} \right\rangle  = \sum\limits_n {{\beta_{n{\bf{k}}}}(t){e^{i{\bf{r}} \cdot {\bf{A}}(t)}}\left| {n,{\bf{k}} - {\bf{A}}(t)} \right\rangle {e^{ - i\int_{ - \infty }^t {{E_{n,{\bf{k}} - {\bf{A}}(t')}}{\rm{d}}t'} }}} 
\end{equation} 
where $\left| {n,{\bf{k}}} \right\rangle $ denotes the eigenstates of the field-free system with band index $n$, and in the position representation has the familiar Bloch form $\left\langle {{\bf{r}}}
\mathrel{\left | {\vphantom {{\bf{r}} {n,{\bf{k}}}}}
	\right. \kern-\nulldelimiterspace}
{{n,{\bf{k}}}} \right\rangle  = \phi _{\bf{k}}^{(n)}({\bf{r}}) = {e^{i{\bf{k}} \cdot {\bf{r}}}}{u_{n{\bf{k}}}}({\bf{r}})$.
In Eq. \ref{Eq:Psi_expansion_ket}, ${{e^{i{\bf{r}} \cdot {\bf{A}}(t)}}\left| {n,{\bf{k}} - {\bf{A}}(t)} \right\rangle {e^{ - i\int_{ - \infty }^t {{E_n}({\bf{k}},t')){\rm{d}}t'} }}}$ denotes the Houston states. Using the ansatz \ref{Eq:Psi_expansion_ket}, Eq. \ref{Eq:TDSE_VG}  reduces to a set of coupled equations among different energy bands for ${{\beta_{n{\bf{k}}}}(t)}$:
\begin{equation}
	\label{Eq:rate_eqn}
	i\frac{d}{{dt}}{\beta _{n{\bf{k}}}}(t) = {\bf{F}}(t) \cdot \sum\limits_m {{\mathbfcal A}_{nm}^{{\bf{k}} - {\bf{A}}(t)}{e^{ - i\int_{ - \infty }^t {\left( {{E_{m,{\bf{k}} - {\bf{A}}(t')}} - {E_{n,{\bf{k}} - {\bf{A}}(t')}}} \right){\rm{d}}t'} }}{\beta _{m{\bf{k}}}}(t)}  
\end{equation} 
where ${\mathbfcal A}_{nm}^{{\bf{k}} - {\bf{A}}(t)} = \left\langle {{u_{n,{\bf{k}} - {\bf{A}}(t)}}\left| {i{\grad _{\bf{k}}}} \right|{u_{m,{\bf{k}} - {\bf{A}}(t)}}} \right\rangle $. where the inner product denotes integration over a unit cell. 
Now let's look into our particular two-band model case, and verify the correspondence between the velocity gauge and the resultant length gauge (Eq. \ref{Eq:Rate_houston}).
In view of the adiabatic theorem, for a driven system if the electron is not allowed to undergo transitions to other bands, then $\psi _{c,v,{\bf{k}}(t)}^{(H)}({\bf{r}}) = \phi _{{\bf{k}}(t)}^{(c,v)}({\bf{r}}){e^{\frac{1}{{i\hbar }}\int_{ - \infty }^t {{E_{c,v}}({\bf{k}}(t')){\rm{d}}t'} }}$ are the instantaneous eigenstates of the time-dependent Hamiltonian, where $\phi _{\bf{k}}^{(c,v)}({\bf{r}}) = {e^{i{\bf{k}} \cdot {\bf{r}}}}{\hat u_{c,v}}({\bf{k}})$ are the Bloch eigenstates for the field-free Hamiltonian. Correspondingly, the most general solution to the time-dependent Schrödinger equation is written as a superposition of adiabatic states:
\begin{equation}
	\label{Eq:Psi_expansion}
	{\Psi _{\bf{k}}}({\bf{r}},t) = \sum\limits_{\bf{k}} {\left[ {{\beta _{v{\bf{k}}(t)}}\psi _{v{\bf{k}}(t)}^{(H)}({\bf{r}}) + {\beta _{c{\bf{k}}(t)}}\psi _{c{\bf{k}}(t)}^{(H)}({\bf{r}})} \right]} \;
\end{equation}
substituting Eq. \ref{Eq:Psi_expansion} into Eq. \ref{Eq:TDSE_modified}
\begin{equation}
	\label{Eq:dummy_1}
	\begin{array}{l}
		i\hbar \left( {\sum\limits_{{\bf{k'}}} {{{\dot \beta }_{v{\bf{k'}}(t)}}\left| {\psi _{v{\bf{k'}}(t)}^{(H)}({\bf{r}})} \right\rangle  + {\beta _{v{\bf{k}}(t)}}\frac{\partial }{{\partial t}}\left| {\psi _{v{\bf{k'}}(t)}^{(H)}({\bf{r}})} \right\rangle  + {{\dot \beta }_{c{\bf{k'}}(t)}}\left| {\psi _{c{\bf{k'}}(t)}^{(H)}({\bf{r}})} \right\rangle  + {\beta _{c{\bf{k}}(t)}}\frac{\partial }{{\partial t}}\left| {\psi _{c{\bf{k'}}(t)}^{(H)}({\bf{r}})} \right\rangle } } \right)\\
		\,\,\,\,\,\,\,\,\,\,\, = \sum\limits_{{\bf{k'}}} {{E_{v{\bf{k'}}}}(t)\left| {\psi _{v{\bf{k'}}(t)}^{(H)}({\bf{r}})} \right\rangle  + {E_{c{\bf{k'}}}}(t)\left| {\psi _{c{\bf{k'}}(t)}^{(H)}({\bf{r}})} \right\rangle } 
	\end{array}
\end{equation}
Multiplying both sides of Eq. \ref{Eq:dummy_1} by  $\left\langle {\psi _{v{\bf{k}}(t)}^{(H)}({\bf{r}})} \right|$ and then integrating by $ \mathbf{r}$, we have 
\begin{equation}
	\label{Eq:rate_beta_v}
	i\hbar \left( {{{\dot \beta }_{v{\bf{k}}(t)}} + {\beta _{v{\bf{k}}(t)}}\left\langle {{{u_{v{\bf{k}}(t)}}}}
		\mathrel{\left | {\vphantom {{{u_{v{\bf{k}}(t)}}} {{{\dot u}_{v{\bf{k}}(t)}}}}}
			\right. \kern-\nulldelimiterspace}
		{{{{\dot u}_{v{\bf{k}}(t)}}}} \right\rangle  + {\beta _{c{\bf{k}}(t)}}\left\langle {{{u_{v{\bf{k}}(t)}}}}
		\mathrel{\left | {\vphantom {{{u_{v{\bf{k}}(t)}}} {{{\dot u}_{c{\bf{k}}(t)}}}}}
			\right. \kern-\nulldelimiterspace}
		{{{{\dot u}_{c{\bf{k}}(t)}}}} \right\rangle {e^{\frac{1}{{i\hbar }}\int_{ - \infty }^t {\left( {{E_{c{\bf{k'}}}}(t') - {E_{v{\bf{k}}}}(t')} \right){\rm{d}}t'} }}} \right) = 0
\end{equation}
where we have taken the orthogonality condition of the Houston basis:
\[\sum\limits_{{\bf{k'}}} {\left\langle {{\psi _{n{\bf{k}}(t)}^{(H)}({\bf{r}})}}
	\mathrel{\left | {\vphantom {{\psi _{n{\bf{k}}(t)}^{(H)}({\bf{r}})} {\psi _{m{\bf{k'}}(t)}^{(H)}({\bf{r}})}}}
		\right. \kern-\nulldelimiterspace}
	{{\psi _{m{\bf{k'}}(t)}^{(H)}({\bf{r}})}} \right\rangle }  = \sum\limits_{{\bf{k'}}} {{e^{\frac{1}{{i\hbar }}\int_{ - \infty }^t {\left( {{E_{v{\bf{k'}}}}(t') - {E_{v{\bf{k}}}}(t')} \right){\rm{d}}t'} }}\left\langle {{{u_{n{\bf{k}}}}}}
	\mathrel{\left | {\vphantom {{{u_{n{\bf{k}}}}} {{u_{m{\bf{k'}}}}}}}
		\right. \kern-\nulldelimiterspace}
	{{{u_{m{\bf{k'}}}}}} \right\rangle \underbrace {\int {d{\bf{r}}{e^{({\bf{k}} - {\bf{k'}}) \cdot {\bf{r}}}}} }_{\delta ({\bf{k}} - {\bf{k'}})}}  = {\delta _{nm}}\delta ({\bf{k}} - {\bf{k'}})\]
as well as the following identities:
\[\left\{ \begin{array}{l}
	\sum\limits_{{\bf{k'}}} {\left\langle {{\psi _{v{\bf{k}}(t)}^{(H)}({\bf{r}})\frac{\partial }{{\partial t}}}}
		\mathrel{\left | {\vphantom {{\psi _{v{\bf{k}}(t)}^{(H)}({\bf{r}})\frac{\partial }{{\partial t}}} {\psi _{v{\bf{k'}}(t)}^{(H)}({\bf{r}})}}}
			\right. \kern-\nulldelimiterspace}
		{{\psi _{v{\bf{k'}}(t)}^{(H)}({\bf{r}})}} \right\rangle }  = \frac{{{E_{v{\bf{k}}(t)}}}}{{i\hbar }} + \sum\limits_{{\bf{k'}}} {{e^{\frac{1}{{i\hbar }}\int_{ - \infty }^t {\left( {{E_{v{\bf{k'}}}}(t') - {E_{v{\bf{k}}}}(t')} \right){\rm{d}}t'} }}\left\langle {{{u_{v{\bf{k}}}}}}
		\mathrel{\left | {\vphantom {{{u_{v{\bf{k}}}}} {{{\dot u}_{v{\bf{k'}}}}}}}
			\right. \kern-\nulldelimiterspace}
		{{{{\dot u}_{v{\bf{k'}}}}}} \right\rangle \underbrace {\int {d{\bf{r}}{e^{({\bf{k}} - {\bf{k'}}) \cdot {\bf{r}}}}} }_{\delta ({\bf{k}} - {\bf{k'}})}}  \\
	\,\,\,\,\,\,\,\,\,\,\,\,\,\,\,\,\,\,\,\,\,\,\,\,\,\,\,\,\,\,\,\,\,\,\,\,\,\,\,\,\,\,\,\,\,\,\,\, = \frac{{{E_{v{\bf{k}}(t)}}}}{{i\hbar }} + \left\langle {{{u_{v{\bf{k}}(t)}}}}
	\mathrel{\left | {\vphantom {{{u_{v{\bf{k}}(t)}}} {{{\dot u}_{v{\bf{k}}(t)}}}}}
		\right. \kern-\nulldelimiterspace}
	{{{{\dot u}_{v{\bf{k}}(t)}}}} \right\rangle \\
	\sum\limits_{{\bf{k'}}} {\left\langle {{\psi _{v{\bf{k}}(t)}^{(H)}({\bf{r}})\frac{\partial }{{\partial t}}}}
		\mathrel{\left | {\vphantom {{\psi _{v{\bf{k}}(t)}^{(H)}({\bf{r}})\frac{\partial }{{\partial t}}} {\psi _{c{\bf{k'}}(t)}^{(H)}({\bf{r}})}}}
			\right. \kern-\nulldelimiterspace}
		{{\psi _{c{\bf{k'}}(t)}^{(H)}({\bf{r}})}} \right\rangle }  = \sum\limits_{{\bf{k'}}} {{e^{\frac{1}{{i\hbar }}\int_{ - \infty }^t {\left( {{E_{c{\bf{k'}}}}(t') - {E_{v{\bf{k}}}}(t')} \right){\rm{d}}t'} }}\left\langle {{{u_{v{\bf{k}}}}}}
		\mathrel{\left | {\vphantom {{{u_{v{\bf{k}}}}} {{{\dot u}_{c{\bf{k'}}}}}}}
			\right. \kern-\nulldelimiterspace}
		{{{{\dot u}_{c{\bf{k'}}}}}} \right\rangle \underbrace {\int {d{\bf{r}}{e^{({\bf{k}} - {\bf{k'}}) \cdot {\bf{r}}}}} }_{\delta ({\bf{k}} - {\bf{k'}})}} \\
	\,\,\,\,\,\,\,\,\,\,\,\,\,\,\,\,\,\,\,\,\,\,\,\,\,\,\,\,\,\,\,\,\,\,\,\,\,\,\,\,\,\,\,\,\,\,\,\, = \left\langle {{{u_{v{\bf{k}}(t)}}}}
	\mathrel{\left | {\vphantom {{{u_{v{\bf{k}}(t)}}} {{{\dot u}_{c{\bf{k}}(t)}}}}}
		\right. \kern-\nulldelimiterspace}
	{{{{\dot u}_{c{\bf{k}}(t)}}}} \right\rangle {e^{\frac{1}{{i\hbar }}\int_{ - \infty }^t {\left( {{E_{c{\bf{k'}}}}(t') - {E_{v{\bf{k}}}}(t')} \right){\rm{d}}t'} }}
\end{array} \right.\]
Eq. \ref{Eq:rate_beta_v} can be further simplified by taking the following ansatz:
\[\begin{array}{l}
	i\hbar \left\langle {{{u_{v{\bf{k}}(t)}}}}
	\mathrel{\left | {\vphantom {{{u_{v{\bf{k}}(t)}}} {{{\dot u}_{v{\bf{k}}(t)}}}}}
		\right. \kern-\nulldelimiterspace}
	{{{{\dot u}_{v{\bf{k}}(t)}}}} \right\rangle  =  - e{\bf{F}}(t) \cdot \left\langle {{u_{v{\bf{k}}(t)}}\left| {i{\grad _{\bf{k}}}} \right|{u_{v{\bf{k}}(t)}}} \right\rangle  =  - e{\bf{F}}(t) \cdot {{\mathbfcal A}_{vv}}({\bf{k}})\\
	i\hbar \left\langle {{{u_{v{\bf{k}}(t)}}}}
	\mathrel{\left | {\vphantom {{{u_{v{\bf{k}}(t)}}} {{{\dot u}_{c{\bf{k}}(t)}}}}}
		\right. \kern-\nulldelimiterspace}
	{{{{\dot u}_{c{\bf{k}}(t)}}}} \right\rangle  =  - e{\bf{F}}(t) \cdot \left\langle {{u_{v{\bf{k}}(t)}}\left| {i{\grad _{\bf{k}}}} \right|{u_{c{\bf{k}}(t)}}} \right\rangle  =  - e{\bf{F}}(t) \cdot {{\mathbfcal A}_{vc}}({\bf{k}})
\end{array}\]
applying the same procedure and multiplying both sides of Eq. (\ref{Eq:dummy_1}) by  $\left\langle {\psi _{c{\bf{k}}(t)}^{(H)}({\bf{r}})} \right|$, respectively, we obtain the following coupled equations for the expansion coefficients:
\begin{equation}
	\label{Eq:coupled_E}
	\left\{ {\begin{array}{*{20}{l}}
			{i\hbar {{\dot \beta }_{v{\bf{k}}(t)}} = e{\bf{F}}(t) \cdot {{\mathbfcal A}_{vv}}({\bf{k}}){\beta _{v{\bf{k}}(t)}} + e{\bf{F}}(t) \cdot {{\mathbfcal A}_{vc}}({\bf{k}}){e^{\frac{1}{{i\hbar }}\int_{ - \infty }^t {\left( {{E_{c{\bf{k'}}}}(t') - {E_{v{\bf{k}}}}(t')} \right){\rm{d}}t'} }}{\beta _{c{\bf{k}}(t)}}}\\
			{i\hbar {{\dot \beta }_{c{\bf{k}}(t)}} = e{\bf{F}}(t) \cdot {{\mathbfcal A}_{cc}}({\bf{k}}){\beta _{c{\bf{k}}(t)}} + e{\bf{F}}(t) \cdot {{\mathbfcal A}_{cv}}({\bf{k}}){e^{ - \frac{1}{{i\hbar }}\int_{ - \infty }^t {\left( {{E_{c{\bf{k'}}}}(t') - {E_{v{\bf{k}}}}(t')} \right){\rm{d}}t'} }}{\beta _{v{\bf{k}}(t)}}}
	\end{array}} \right.
\end{equation}
if we define the Houston functions in terms of the global energy of ${n^{\rm{th}}}$-band, i.e.,  ${E_{n{\bf{k}}}}(t) \to E_{n{\bf{k}}}^T(t) = {E_{n{\bf{k}}}}(t) - e{\bf{F}}(t) \cdot {{\mathbfcal A}_{nn}}({\bf{k}})$, the coupled equations (\ref{Eq:coupled_E}) reduces analogously to Eq. \ref{Eq:Rate_houston}.\

\section*{Data availability} The data that support the findings of this study are available from the authors upon reasonable request.

%

\bibliography{article}

\end{document}